\documentclass[conference]{IEEEtran}
\IEEEoverridecommandlockouts

\usepackage{cite}
\usepackage{verbatim}
\usepackage{amsmath,amssymb,amsfonts}
\usepackage{algorithmic}
\usepackage{graphicx}
\usepackage{textcomp}
\usepackage{xcolor}
\usepackage{amsmath}
\usepackage{caption}
\usepackage{subcaption}
\usepackage{mdframed,steinmetz}
\usepackage{booktabs} 
\usepackage{siunitx}  

\renewcommand{\theenumi}{\bf\arabic{enumi}}
\renewcommand{\theenumii}{\bf\alph{enumii}}

\newcommand{\phat}{{{\widehat{p}}}}

\newcommand{\bx}{{{\bf x}}}

\newcommand{\bzhat}{{\widehat{\bf z}}}

\newcommand{\thetahat}{{\widehat {\theta}}}
\newcommand{\EQ}{\begin{equation}}
\newcommand{\EN}{\end{equation}}
\newcommand{\ben}{\begin{enumerate}}
\newcommand{\een}{\end{enumerate}}

    \def\squarebox#1{\hbox to #1{\hfill\vbox to #1{\vfill}}}

\renewcommand{\thefootnote}{}
 
\mathchardef\inn="3232
\renewcommand{\in}{\mbox{$\,\inn\,$}}

\def\BibTeX{{\rm B\kern-.05em{\sc i\kern-.025em b}\kern-.08em
    T\kern-.1667em\lower.7ex\hbox{E}\kern-.125emX}}

\usepackage{stfloats}
\graphicspath{ {./Figs/} }

\begin{document}

\title{mmSnap: Bayesian One-Shot Fusion in a Self-Calibrated mmWave Radar Network\\
\thanks{This work was supported in part by the Center for Ubiquitous Connectivity (CUbiC), sponsored by Semiconductor Research Corporation (SRC) and Defense Advanced Research Projects Agency (DARPA) under the JUMP 2.0 program, and in part by the National Science Foundation under grant CNS-2215646.}
}

\author{\IEEEauthorblockN{Anirban Banik\IEEEauthorrefmark{2}, Lalitha Giridhar\IEEEauthorrefmark{2}, Aaditya Prakash Kattekola\IEEEauthorrefmark{2}, Anurag Pallaprolu\IEEEauthorrefmark{2},\\
Yasamin Mostofi\IEEEauthorrefmark{2}, Ashutosh Sabharwal\IEEEauthorrefmark{4} and Upamanyu Madhow\IEEEauthorrefmark{2}}
\IEEEauthorblockA{
\IEEEauthorrefmark{2}University of California, Santa Barbara, CA, U.S.A. \; \IEEEauthorrefmark{4}Rice University, TX, U.S.A}
\{abanik, lalitha, aadityaprakash, apallaprolu, ymostofi, madhow\}@ucsb.edu \; ashu@rice.edu}

\maketitle

\sisetup{
  table-format = 2.4,    
  detect-mode,
  detect-family,
  detect-weight
}

\begin{abstract}
We present \textit{mmSnap}, a collaborative RF sensing framework using multiple radar nodes, and demonstrate its feasibility and efficacy using commercially available mmWave MIMO radars.  Collaborative fusion requires network calibration, or estimates of the relative poses (positions and orientations) of the sensors. We experimentally validate a self-calibration algorithm developed in our prior work, which estimates relative poses in closed form by least squares matching of target tracks within the common field of view (FoV). 
\noindent  We then develop and demonstrate a Bayesian framework for one-shot fusion of measurements from multiple calibrated nodes, which yields instantaneous estimates of position and velocity vectors that match smoothed estimates from multi-frame tracking. Our experiments, conducted outdoors with two radar nodes tracking a moving human target, validate the core assumptions required to develop a broader set of capabilities for networked sensing with opportunistically deployed nodes.
\end{abstract}
\section{Introduction}
\begin{figure}[t!] 
 \centering 
\includegraphics[scale = 0.35]{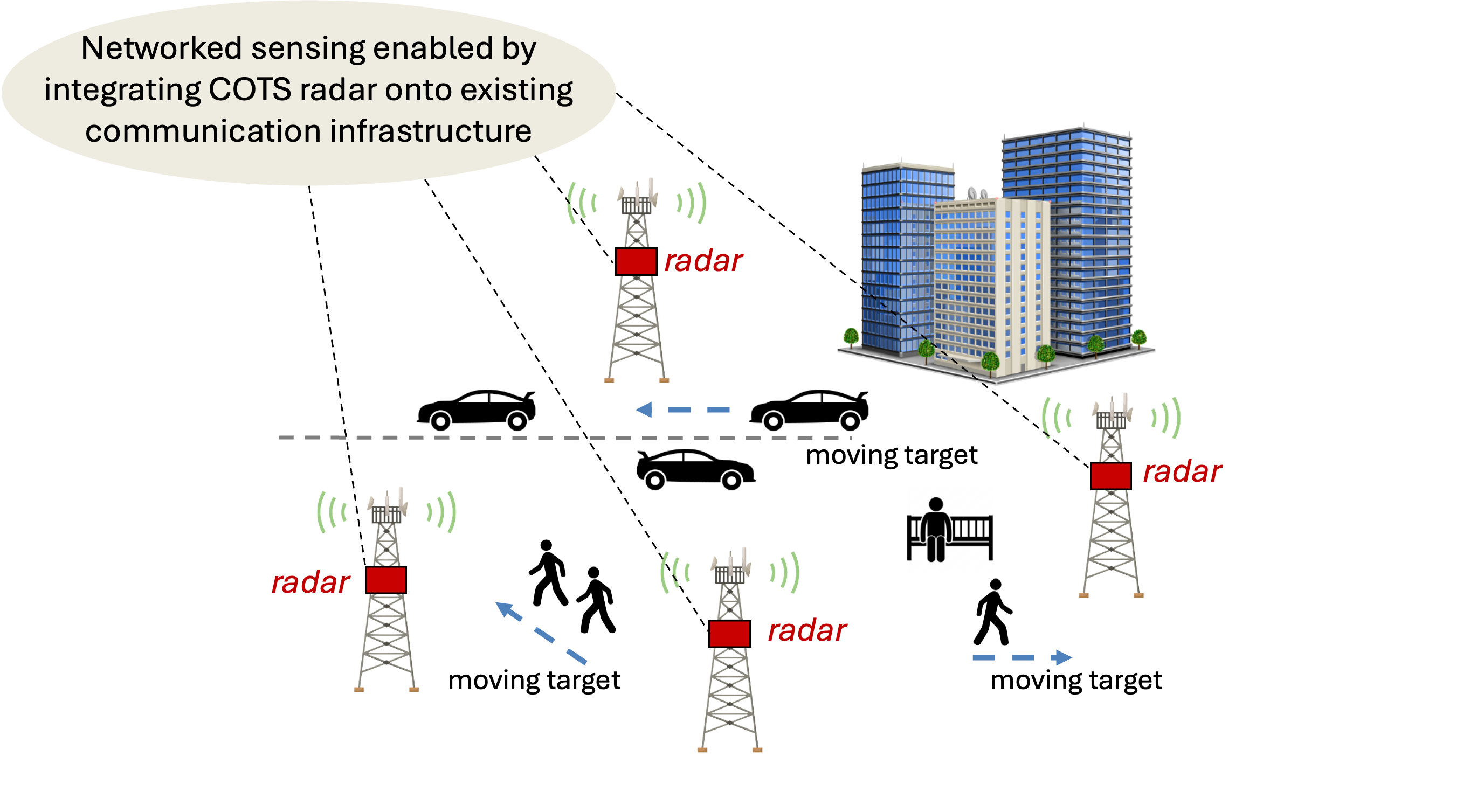} 
\caption{A collaborative networked RF sensing infrastructure.}
\label{fig:distRadar}
\vspace{-0.6cm}
\end{figure}

The integration of low-cost millimeter-wave (mmWave) MIMO radar in the 60 and 77 GHz bands with a communication infrastructure (e.g., cellular or WiFi) enables the creation of collaborative RF sensing networks, as illustrated in Figure~\ref{fig:distRadar}. This approach can overcome the coverage limitations of individual radar nodes, which face line of sight (LoS) obstructions and limited field of view (FoV). When radar nodes with overlapping FoVs collaborate, they can significantly enhance the quality of inference by fusing their measurements. For instance, a single mmWave radar node can provide a precise estimate of Doppler (radial velocity), which is the projection of a moving target's velocity along the LoS. However, it cannot instantaneously estimate the target’s full vector velocity. By combining Doppler measurements from two radar nodes, the vector velocity can be estimated instantaneously, except in degenerate configurations. Similarly, individual nodes can make accurate position estimates by combining high-resolution range data (due to the large bandwidth available) with angle measurements, though the latter is limited by the aperture size of the radar. Collaborative sensing allows multiple nodes to combine their range estimates, enhancing the resolution of position estimates. Such fine-grained fusion relies on precise network calibration, requiring accurate estimates of the nodes' relative positions and orientations (poses). In opportunistically deployed networks, where such information is typically unavailable, calibration becomes a major challenge but remains essential for effective collaborative sensing.

\noindent \textbf{Contributions:} We introduce \textit{mmSnap}, a framework for self-calibration and Bayesian one-shot fusion in a radar network, designed to deliver instantaneous (``snap'') position and velocity through fusion in a millimeter-wave (mmWave) radar network.  Self-calibration is based on an optimal algorithm for joint tracking and pose estimation introduced in our prior work \cite{10942741}, in which closed form estimates for the relative poses for any two nodes tracking a target in their shared FoV are obtained by least squares alignment of the estimated tracks.  Then, assuming a calibrated network, we develop a Bayesian framework for estimating instantaneous target position and velocity based on measurements from collaborating nodes. We present preliminary experimental results using COTS 77 GHz mmWave radar nodes in an outdoor setting, validating that the self-calibration algorithm from \cite{10942741} yields accurate pose estimates from human target tracks, and that post-calibration, our Bayesian one-shot fusion method produces instantaneous estimates close to smoothed multi-frame tracking outputs. While these initial experiments are limited to two radar nodes and a single moving target, they establish the feasibility of core building blocks for collaborative RF sensing at scale.

\begin{figure}[t!] 
 \centering 
\includegraphics[scale = 0.26]{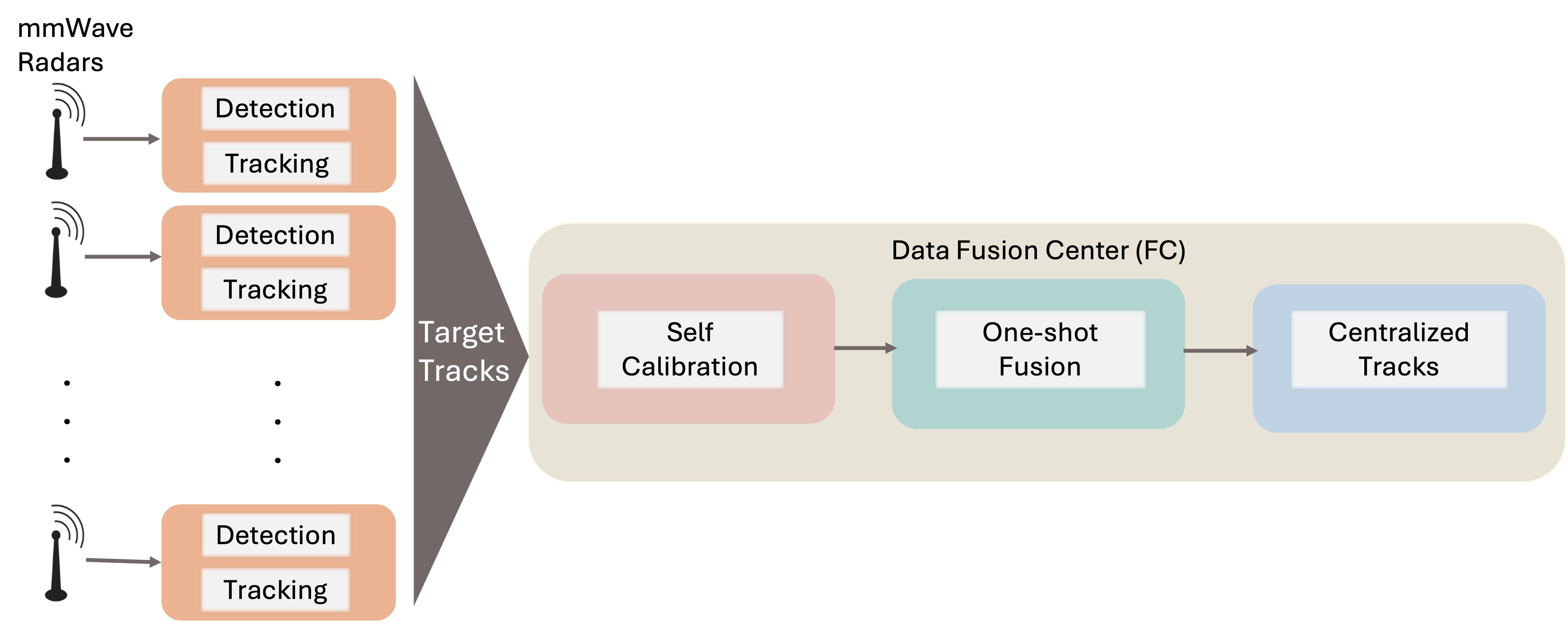} 
\caption{The end-to-end mmSnap pipeline. }
\label{fig:mmSnapPipeline}
\vspace{-0.6cm}
\end{figure}

\noindent \textbf{Related work:} 
The use of distributed radar for estimating vector velocity has been explored in several studies~\cite{6856449,9906497,10793274,6018296}: ~\cite{6856449} assesses single-shot ego-velocity for vehicles, while~\cite{9906497} uses a linear least-squares method for velocity estimation amidst Doppler ambiguities. In~\cite{10793274}, Doppler and angular data from two radars are fused, and~\cite{6018296} proposes a maximum likelihood method, for velocity estimation which is prone to struggle in noisy conditions. Unlike our framework, which includes self-calibration and Bayesian fusion, these studies assume known radar poses relative to a global frame, which may be unrealistic. Further, the maximum likelihood techniques that they employ struggle in the presence of geometric degeneracy.  A more general problem of scene estimation and point cloud based environmental mapping with multiple radars is considered in~\cite{10577265}, with Doppler information utilized for registration across radars.  Our approach is computationally much simpler because it focuses on self-calibration and one-shot fusion for point target models, demonstrating the efficacy of approximating extended human targets as detection-cluster centroids for both tasks. Finally, recent efforts in self-calibration using a target in the common FoV include brute force optimization for 3D pose estimation using a corner reflector target in~\cite{9924790}, and singular value decomposition (SVD) for 2D pose estimation using human targets in~\cite{9764173}.  Our self-calibration approach is similar to the latter, but we utilize the complex number framework in our prior work~\cite{10942741}, which yields relative poses in closed form. 

To the best of our knowledge, this is the first work to experimentally demonstrate a framework for self-calibration and Bayesian one-shot fusion in radar networks geared toward urban outdoor settings. While our approach is applicable to general moving objects, we validate the framework through controlled experiments with human motion. Results across varied radar configurations show that our approach is well-suited to collaborative sensing with opportunistic node deployment, where significant variation in the relative poses of nodes with overlapping FoVs is expected.

The remainder of this paper is structured as follows: Section~\ref{sec:mmSnapPipeline} presents the full mmSnap system model and self-calibration algorithm; Section~\ref{sec:oneshot} details the Bayesian framework for one-shot fusion. Experimental results for self-calibration and fusion in a two-radar network are provided in each of these sections. Section~\ref{sec:conclusion} offers concluding remarks and outlines future directions.

\section{mmSnap System Model}\label{sec:mmSnapPipeline}

\begin{figure*}[t!] 
 \centering 
\includegraphics[scale = 0.45]{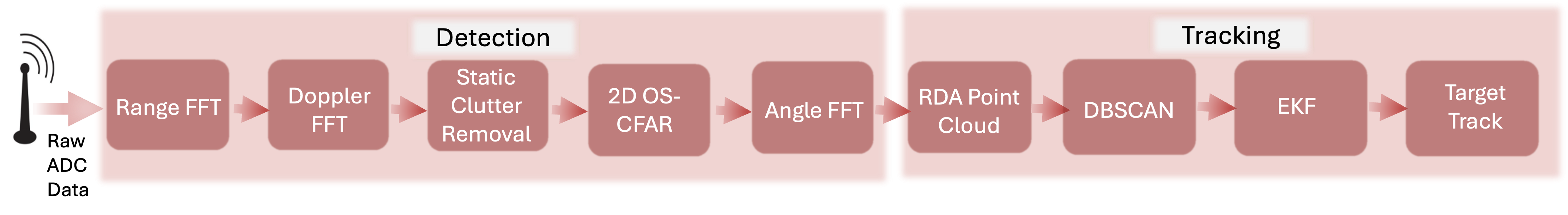} 
\caption{mmSnap Signal Processing Chain from Raw ADC data to Target Tracks.}
\label{fig:mmSnapPipeline_sp}
\vspace{-0.6cm}
\end{figure*}
A possible realization of the end-to-end mmSnap pipeline is illustrated in Figure \ref{fig:mmSnapPipeline}.  Each radar node carries out conventional point cloud detection individually. Depending on the available computation, it may also perform target tracking on its own, as shown in Figure \ref{fig:mmSnapPipeline}, or this functionality could be migrated to the fusion center.   Each node shares its detection and tracking (if performed at the node) results with the fusion center.  Track matching for common targets is used for self-calibration. One-shot fusion of detected points is used to provide low-latency estimates of target position/velocity, while also serving as inputs to a centralized tracker. The nodes are coarsely time-synchronized at the level of a radar frame.

Our experiments employ 77GHz AWR2243 MIMO radar boards from Texas Instruments, with each board outputting raw ADC data in radar cube format. The chirp parameters used in the experiments, and the corresponding range and radial velocity resolutions, are detailed in Table~\ref{tab:chirp_parameters}. 
The radar boards are equipped with three transmit and four receive antennas, resulting in 12 virtual antennas with an inter-antenna spacing of \(d = \lambda/2\). However, the virtual array size for azimuth angle estimation (which is what we focus on) is 8, corresponding to a spatial frequency resolution of $2\pi/8=\pi/4$. While the mmSnap architecture in Fig.~\ref{fig:mmSnapPipeline} assumes real-time communication with a fusion center, our experiments use offline processing of data collected from multiple sensors.
\subsection{Signal Processing Pipeline}\label{sec:mmsnapsp}

As illustrated in Fig.~\ref{fig:mmSnapPipeline_sp}, the local radar signal processing at each node is standard. The raw ADC data is transformed from its original radar cube format into Range-Doppler-Angle point clouds for each frame by applying Fast Fourier Transforms (FFTs) along the corresponding dimensions. To focus on moving targets, static clutter removal is performed, followed by a two-dimensional Ordered-Statistics Constant False Alarm Rate (2D OS-CFAR) detection applied across frames for further refinement. In the tracking stage, DBSCAN clustering extracts point cloud centroids, which are then processed by an Extended Kalman Filter (EKF) for continuous tracking. For human targets within 10 m, the point cloud shows significant Doppler variation due to limb movements. We design a DBSCAN variant to extract a cluster center representing torso position and motion, allowing the self-calibration and one-shot fusion algorithms, based on point target models, to remain simple and effective.

\begin{table}[t!]  
    \centering
    \renewcommand{\arraystretch}{1.3}
    \begin{tabular}{l c}
        \toprule
        \textbf{Parameter} & \textbf{Value} \\
        \midrule
        Carrier Frequency ($f_c$) & 76 GHz \\
        ADC Sampling Rate & 10 MHz \\
        Number of Transmitters & 3 \\
        Number of Receivers & 4 \\
        Bandwidth & 4.24 GHz \\
        Chirp Slope & 83 MHz/$\mu$s \\
        Frame Duration & 150 ms \\
        Number of Chirps per Frame & 60 \\
        Chirp Ramp Time & 60 $\mu$s \\
        Range Resolution & 3.54 cm \\
        Doppler Resolution & 0.183 m/s \\
        Max Unambiguous Range & 18.07 m \\
        \bottomrule
    \end{tabular}
    \caption{Radar Parameters for mmSnap Experiments}
    \label{tab:chirp_parameters}
\end{table}

\captionsetup[subfloat]{labelformat=simple}
\renewcommand{\thesubfigure}{(\alph{subfigure})}

\begin{figure}[t!]
   \centering
   \subfloat[]{\includegraphics[width=0.2\textwidth]{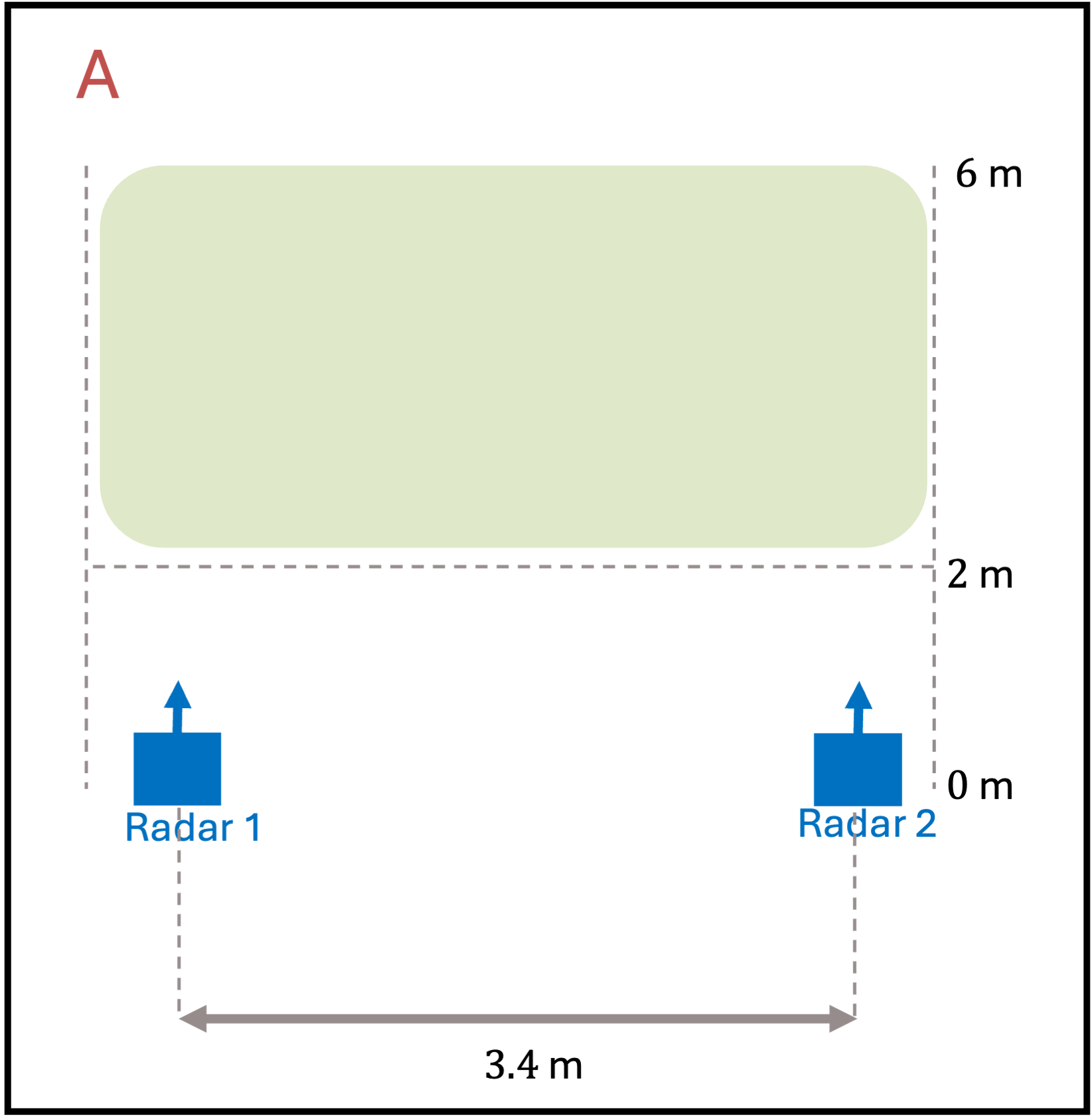}\label{a1}}
   \hspace{0.5em}
   \subfloat[]{\includegraphics[width=0.2\textwidth]{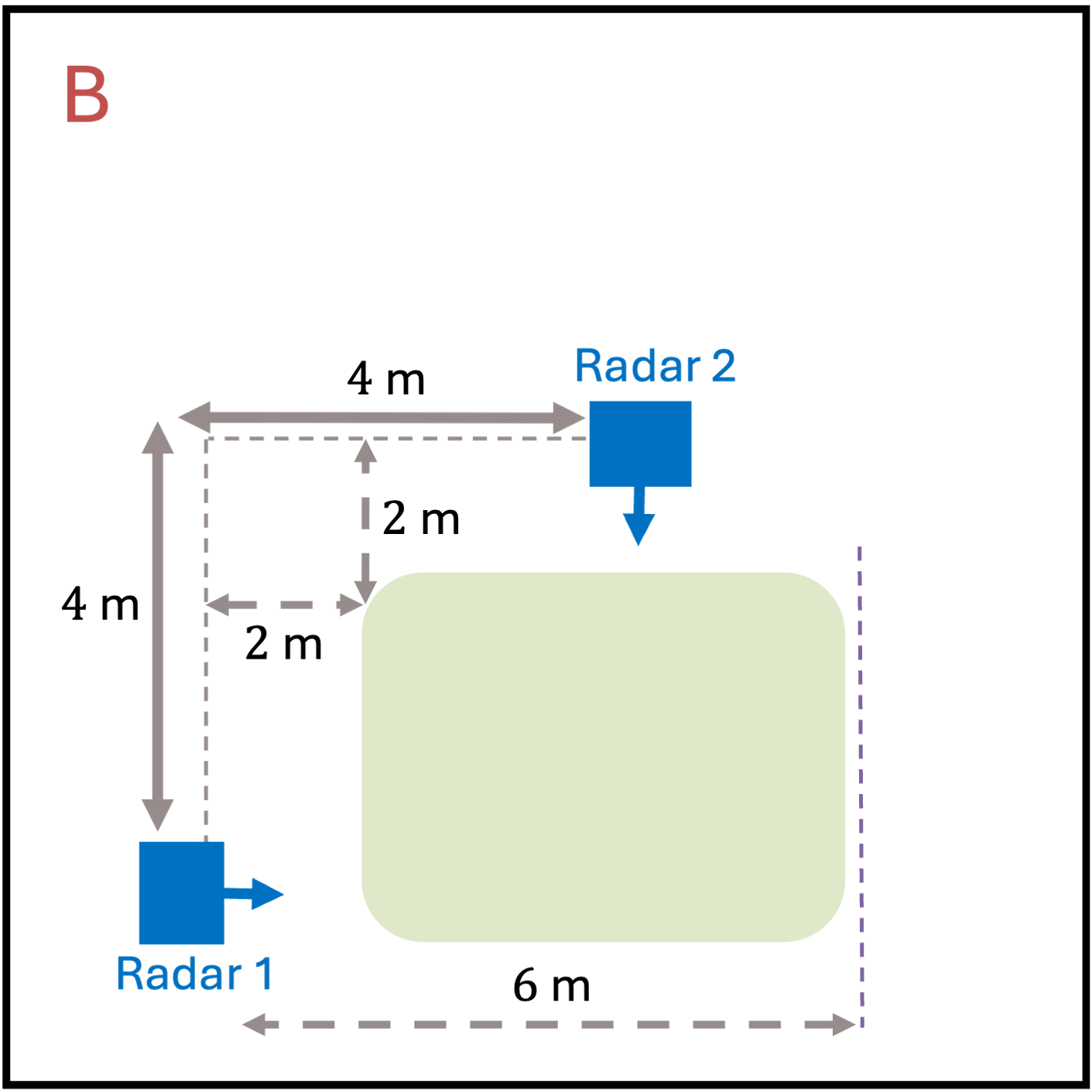}\label{a2}}

   \vspace{1em}

   \par\medskip
   \subfloat[]{\includegraphics[width=0.2\textwidth]{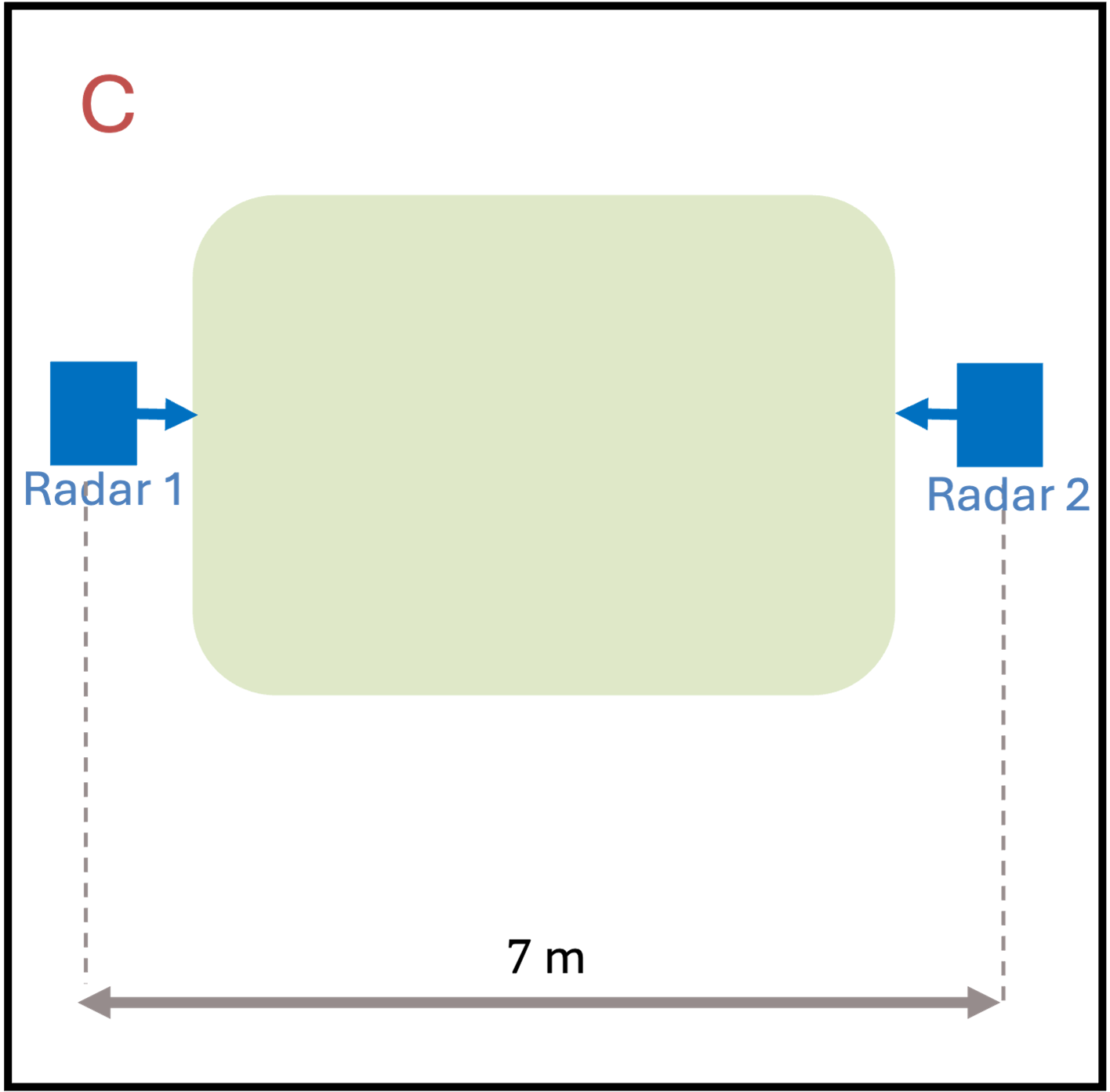}\label{b1}}

   \caption{Illustration of the experimental setups used to evaluate the mmSnap pipeline, labeled A through C. The green regions demarcate the area within which the target moves, following two different trajectories: either a straight line or random motion.} 
   \label{ExpSettings}
\end{figure}

We consider a two-radar outdoor network tracking a single moving target, following either a straight line or a random trajectory.  Three two-node configurations (A-C in Fig.\ref{ExpSettings}), differing in the relative orientation of the arrays in the nodes, are investigated. The array broadside for each node is indicated by an arrow, and the shaded region represents the area within which the target moves.

\subsection{Measurement Model}\label{sec:measurementmodel}
\begin{figure}[t!] 
 \centering 
\includegraphics[scale = 0.5]{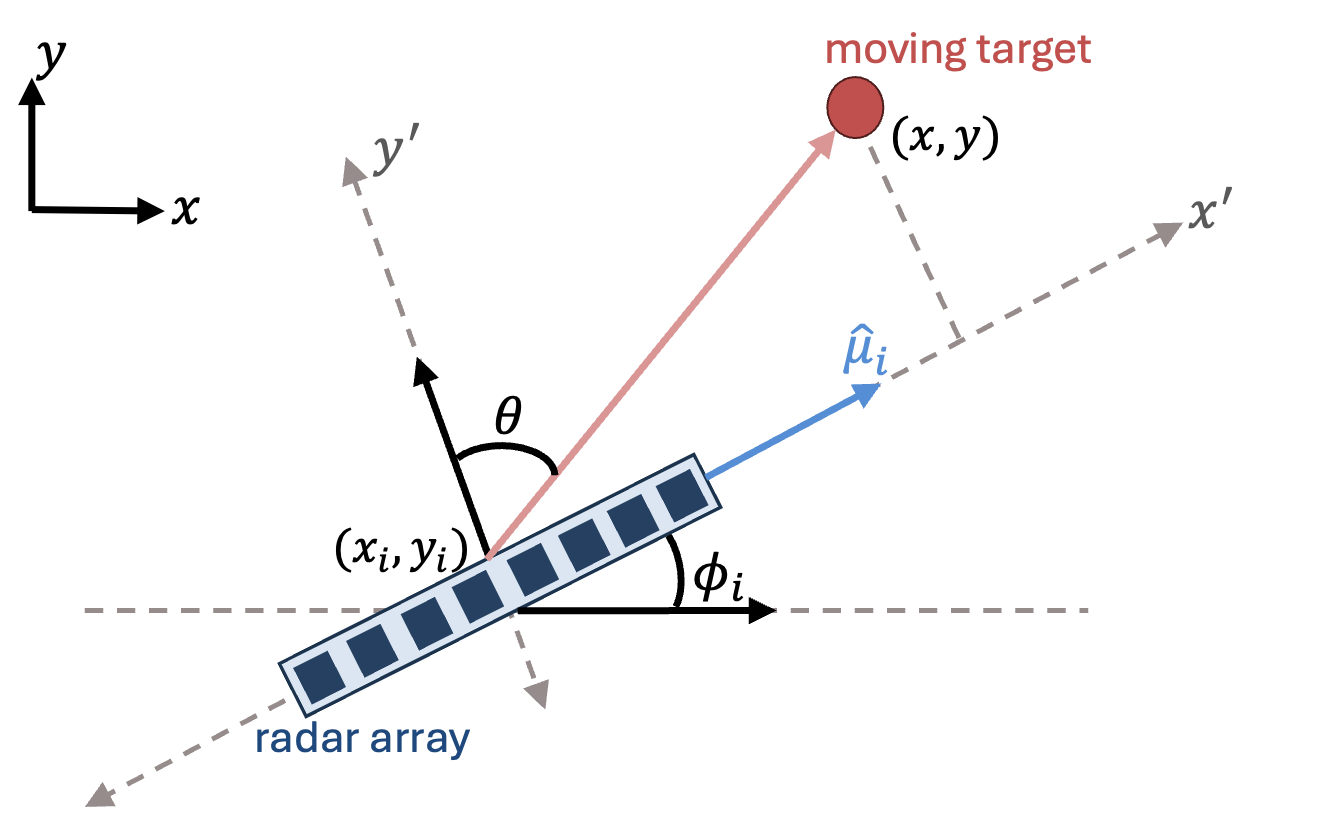} 
\caption{Measurement model for a single moving target seen by a radar with respect to a global frame of reference}
\label{fig:measurementmodel}
\vspace{-0.6cm}
\end{figure}

We begin by introducing the basic measurement model for a radar node $S_i$, where $i \in {1, \dots, N}$ in an $N$-node network, tracking a single moving target, as illustrated in Fig.~\ref{fig:measurementmodel}. 
As shown in Fig.~\ref{fig:measurementmodel}, the positions of radar node $S_i$ and the target with respect to a global reference frame are denoted by 2D vectors $\mathbf{p_i} = (x_i, y_i)^T$ and $\mathbf{p} = (x, y)^T$, respectively, and the target velocity is denoted by $\mathbf{\dot{p}} = (v_x, v_y)^T$. 
The target range is given by
\EQ \label{range_estimate}
    r_{i} = \| \mathbf{p} - \mathbf{p_i} \| = \sqrt{(x - x_{i})^{2} + (y - y_{i})^{2}}
\EN
The radial velocity measured via Doppler is obtained by projecting the target velocity vector onto the unit vector pointing from the radar node to the target:
\EQ \label{Doppler_estimate}
    v_{i} = \bigg \langle {\mathbf{\dot{p}}, \frac{\mathbf{p}-\mathbf{p_i}}{\| \mathbf{p} - \mathbf{p_i} \|}} \bigg \rangle
    = \frac{v_{x}\,(x - x_{i}) + v_{y}\,(y - y_{i})}{r_{i}}
\EN
where $\langle \mathbf{a}, \mathbf{b} \rangle = \mathbf{a}^T \mathbf{b}$ denotes the inner product of vectors $\mathbf{a}$ and $\mathbf{b}$. 
For a $\lambda/2$-spaced virtual array, the spatial frequency of the target
is given by 
\EQ \label{spatial_frequency_estimate}
\omega_{i} = \pi \sin{\theta} = \frac{\pi\langle {\mathbf{p} - \mathbf{p_i}, \mathbf{\hat{\mu_i}}} \rangle}{r_i}
=\frac{((x-x_i)\cos{\phi_i}+(y-y_i)\sin{\phi_i})}{r_i}
\EN
where $\mathbf{\hat{\mu_i}} = (\cos{\phi_i}, \sin{\phi_i})^T$ denote the unit vector along the radar array. In order to apply the preceding measurement model across radar nodes, we need a global frame of reference, which is established via self-calibration, as described next.

\subsection{Self Calibration}
Self-calibration is accomplished by matching the ``local'' tracks seen by different nodes for a target in their common FoV. We evaluate pairwise self-calibration in our 2-node experiments, but as noted in \cite{10942741}, these can be stitched together for network-wide calibration. Expressing 2D coordinates as complex numbers, denote by $z_i[k] = u_i[k] + j v_i[k]$, $i=1,2$, the target's position in the $k$th radar frame as estimated in node $S_i$'s coordinate system.
Letting $S_1$ serve as the global reference without loss of generality, the optimal estimate for the relative (complex-valued) position \( p_{21} \) and relative orientation \( \phi_{21} \in [0, 2\pi) \) of \( S_2 \) with respect to \( S_1 \) is obtained by least squares minimization of the track matching error:
\EQ \label{least_squares2}
J(p, \phi ) = \sum_{k=1}^K \big{|} z_1 [k] - p  - e^{j \phi} z_2 [k] \big{|}^2
\EN
The solution is obtained in closed form as \cite{10942741}:
\EQ \label{two_radar_calibration}
\begin{array}{l}
\hat{\phi}_{21} = -\phase{\bzhat_1^H \bzhat_2}\\
 \\
\phat_{21} = \overline{z}_1 - e^{j  \hat{\phi}_{21}} \overline{z}_2
 \end{array}
\EN
where \( \overline{z}_i \) is the track centroid, \( \bzhat_i \) is the centered track from \( S_i \)'s perspective, defined as \( \bzhat_i = (z_i[1] - \overline{z}_i, \dots, z_i[K] - \overline{z}_i)^T \), and 
\( \bx^H = (\bx^*)^T \) denotes the conjugate transpose of a vector \( \bx \). The minimum cost achieved by this solution is given by
\EQ \label{min_cost}
J_{min} (2,1)  = ||\bzhat_1||^2 + ||\bzhat_2||^2 - 2 \big{|} \bzhat_1^H \bzhat_2 \big{|}
\EN
The trajectory root-mean-square error (RMSE) after self-calibration is given by
\EQ \label{trajectory_rmse}
RMSE = \sqrt{J_{min}/K}
\EN
The rationale for self-calibration is the difficulty of estimating the ground truth for relative poses in large-scale opportunistic deployments. Consequently, the track fitting RMSE serves as a useful, albeit indirect, quantitative indicator of calibration accuracy.
\begin{table}[t]
  \centering
  \renewcommand{\arraystretch}{1.3}

  \resizebox{\columnwidth}{!}{%
    \begin{tabular}{@{}l S S c S S c S S@{}}
      \toprule
      & \multicolumn{1}{c}{$A$} & \phantom{abc}& \multicolumn{1}{c}{$B$} &
        \phantom{abc} & \multicolumn{1}{c}{$C$} \\
      \midrule
      \multicolumn{1}{l}{\textbf{RMSE (m)}}\\
      Straight Trajectory & 0.5664 && 0.5657 && 0.1783 \\
      Random Trajectory & 0.7076 && 0.6331 && 0.4426 \\
      \bottomrule
    \end{tabular}
  }%

  \caption{Trajectory RMSE values for different self-calibration scenarios}
  \label{tab:residual}
\end{table}

\begin{figure}[t!]
    \centering
    \subfloat[]{\includegraphics[width=0.48\columnwidth]{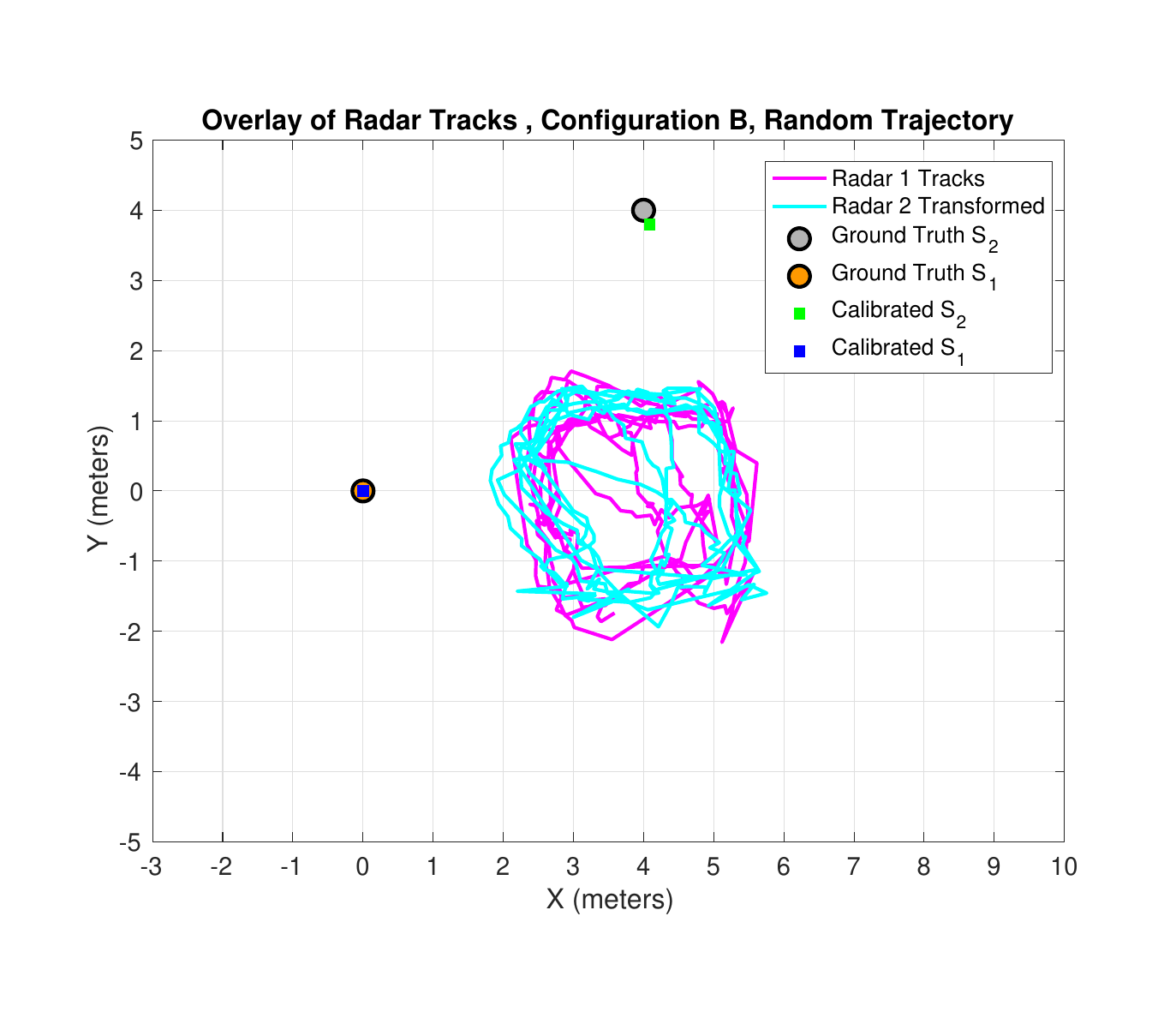}\label{b2}} 
    \vspace{0.3em}
    \subfloat[]{\includegraphics[width=0.48\columnwidth]{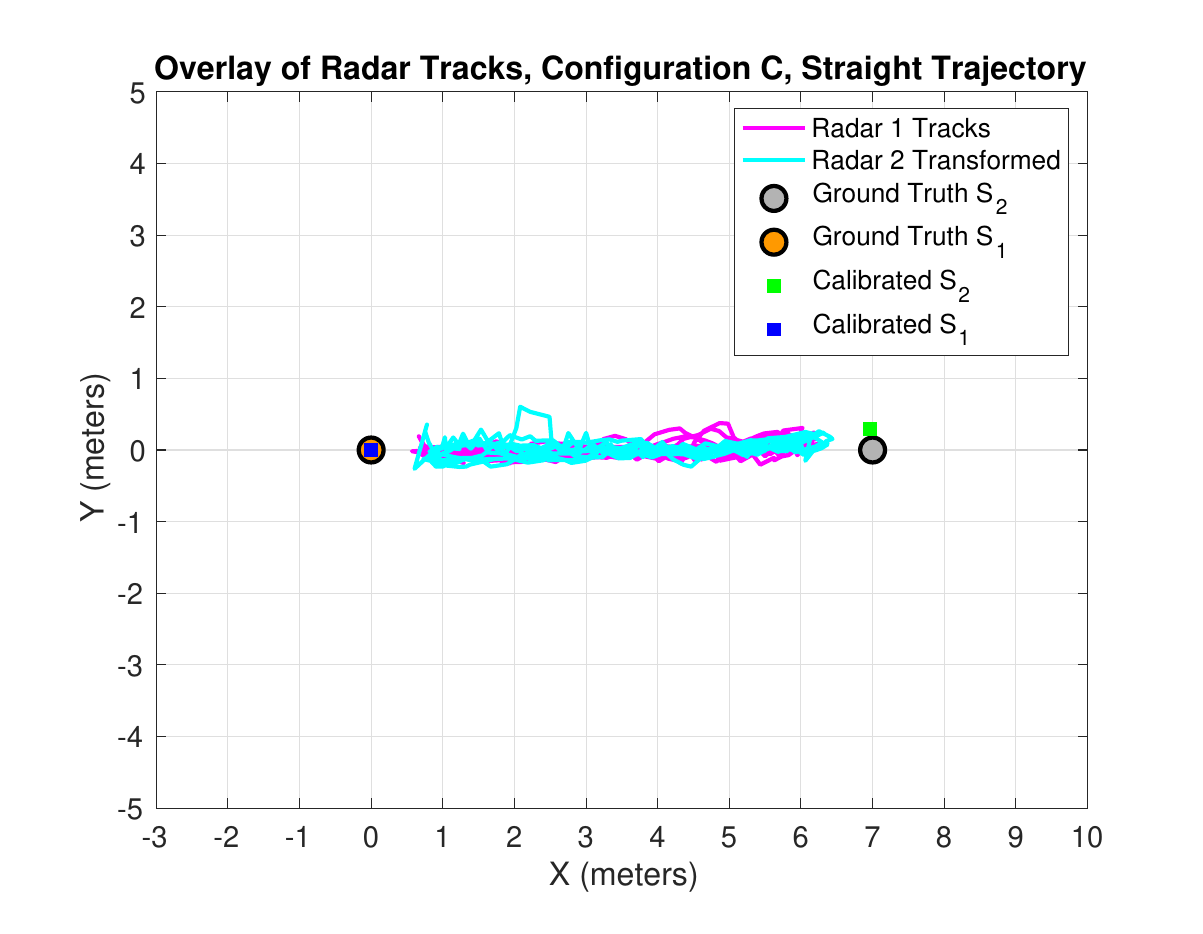}\label{c1}}
    \caption{Overlaid tracks after self-calibration for radar node configurations B and C, with random and straight target trajectories, respectively.}
    \label{fig:tracks_overlayed}
\end{figure} 

For experimental validation of self-calibration, we collect $90$s sequences of a single moving human target within the overlapping FoV of the two radars. For radar frame duration of $150$ ms, each sequence yields a trajectory of length $K=600$ frames for self-calibration. We show in Table~\ref{tab:residual} the RMSE as computed using (\ref{trajectory_rmse}) for different node configurations (Fig.~\ref{ExpSettings}) and trajectories (straight line and random). While low RMSE indicates good track alignment and hence acceptable accuracy in relative poses, it is not a direct measure of pose accuracy, since it also depends on node configuration and the nature of the trajectory. For instance, straight-line trajectories yield lower RMSE across configurations, yet pose estimates remain comparable to those from random trajectories. In setup C, the estimates are \((0.29\,\text{m},\,6.96\,\text{m},\,184.4^\circ)\) for the straight and \((0.03\,\text{m},\,6.90\,\text{m},\,178.6^\circ)\) for the random trajectory, both close to the ground truth of \((0\,\text{m},\,7\,\text{m},\,180^\circ)\). In Fig.~\ref{fig:tracks_overlayed}, we visually demonstrate the self-calibration algorithm by overlaying the track seen by $S_1$ and that seen by $S_2$, after transforming the latter to $S_1$'s frame based on its estimated relative pose.  Since the tracking by each radar is imperfect, the overlay is also noisy. However, averaging across multiple frames yields accurate estimates of the relative pose. 

\section{One-shot Fusion}\label{sec:oneshot}

We now extend the measurement model from Section~\ref{sec:measurementmodel} to a multi-radar setting, derive the Bayesian one-shot fusion framework, and present experimental comparisons with smoothed multi-frame estimates.

\subsection{Bayesian Framework}
Consider $N$ radars $S_i, i=1, ..., N$ tracking a common target moving in their overlapping FoV. We assume that the network is calibrated (i.e., relative poses are known), and without loss of generality, set node $S_1$ as our reference, so that $\mathbf{p_1}=(0,0)^T, \phi_1=0^\circ$. 
The range ($r_i$), radial velocity ($v_i$) and spatial frequency ($\omega_i$) estimates for $S_i, i=1, ..., N$, are then given by (\ref{range_estimate})-(\ref{spatial_frequency_estimate}).  The corresponding measurements are modeled as conditional Gaussian as follows
\EQ \label {Meas Error Distributions}
\begin{array}{l}
     R_{i} \sim \mathcal{N}(r_i, \sigma_{R}^2)\\
     \Omega_{i} \sim \mathcal{N}(\omega_i, \sigma_{\Omega}^2)\\
     V_{i} \sim \mathcal{N}(v_i, \sigma_{V}^2)\\
     
\end{array}
, \forall  i \in \{1, 2, ..., N\}
\EN
where $\sigma_{R}^2$, $\sigma_{\Omega}^2$, and $\sigma_{V}^2$ are the measurement noise variances in range, spatial frequency, and radial velocity, respectively. The target state $\theta$ is taken to be the instantaneous position and velocity with respect to the global frame, $\theta = (x, y, v_x, v_y)^T$. For one-shot fusion, we estimate the state based on the observation vector $\mathcal{Y}$ corresponding to the actual range, radial Doppler and spatial frequency measurements of the target by the $N$ radars
\[
\mathcal{Y}=(R_1, R_2, ..., R_N, \Omega_1, \Omega_2, ..., \Omega_N, V_{1}, V_{2}, ..., V_N)^T
\]
Assuming that the measurement noises in  (\ref{Meas Error Distributions}) are independent across sensors and across modalities, we model $\mathcal{Y}$ is conditionally Gaussian given the state $\theta$. 

The maximum likelihood (ML) estimate of $\theta$,
\EQ \label {ML}
\thetahat_{ML} = {\rm arg~max_{\theta}}~p(\mathcal{Y} \mid \theta)
\EN
is given by minimizing the negative log likelihood $L_{ML}(\theta)=-\log p(\mathcal{Y} \mid \theta)$, and is the solution to the following nonlinear least squares problem
\EQ \label {ML Estimate}
\begin{array}{cc}
\thetahat_{ML} = {\rm arg~min_{\theta}}~L_{ML}(\theta)\\\\
= {\rm arg~min_{\theta}}~[\frac{1}{\sigma_{R}^2}\sum_{i=1}^{N}(R_i-r_i)^2 + \frac{1}{\sigma_{\Omega}^2}\sum_{i=1}^{N}(\Omega_i-\omega_i)^2 \\
+ \frac{1}{\sigma_{V}^2}\sum_{i=1}^{N}(V_{i}-v_{i})^2]
\end{array}
\EN

\begin{figure}[t!] 
 \centering 
\includegraphics[scale = 0.43]{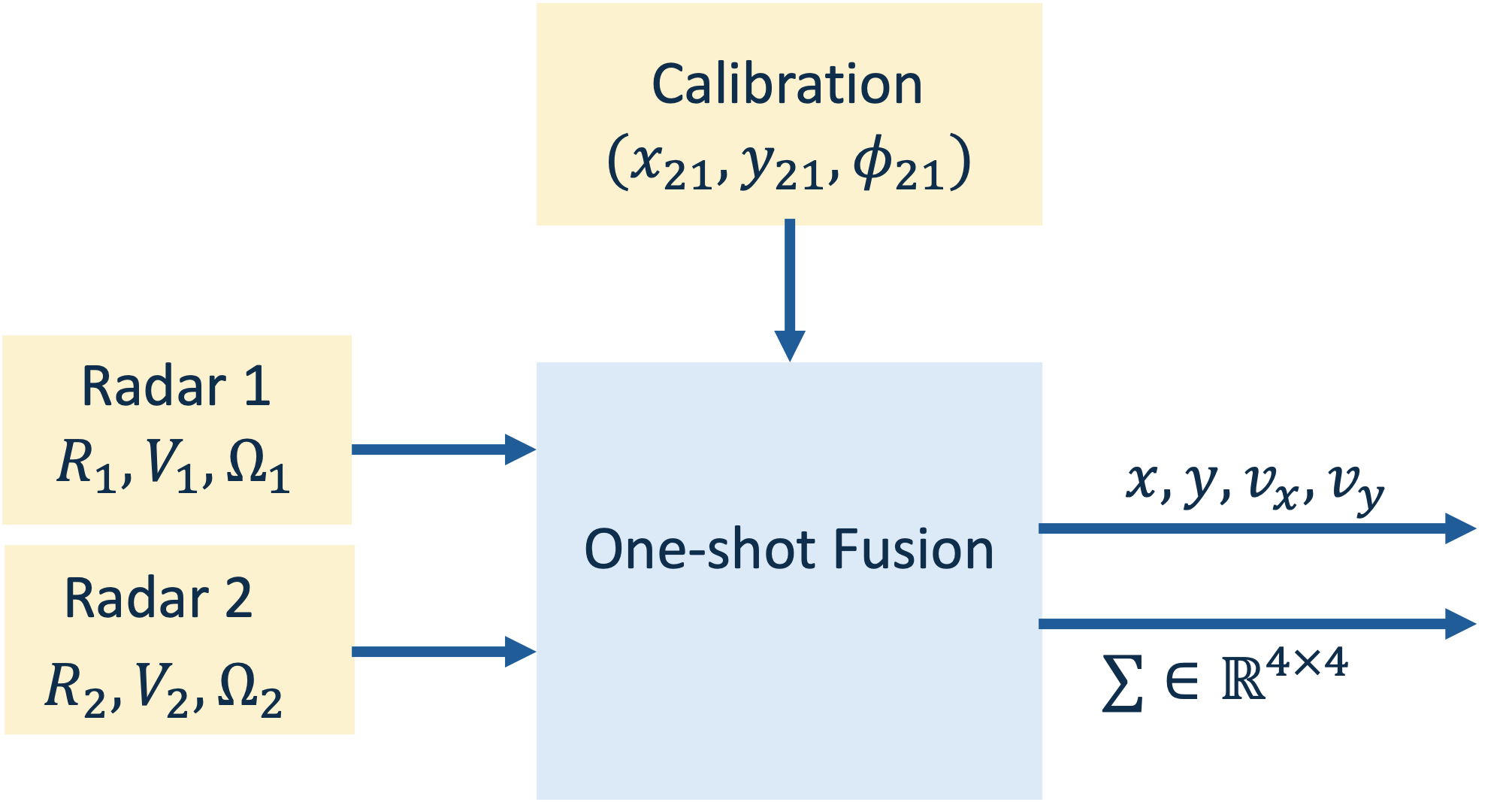} 
\caption{The one-shot fusion pipeline for a two-radar network.}
\label{fig:oneShotPipeline}
\vspace{-0.6cm}
\end{figure}

\noindent \textit{Limitation:} The ML one-shot estimate is usually effective, but can yield unstable estimates in degenerate settings, as discussed later.  We therefore regularize the nonlinear least squares problem in (\ref{ML Estimate}) using a Bayesian framework using a Gaussian prior for the state:
\EQ \label {Prior Distributions}
\begin{array}{l}
     x \sim \mathcal{N}(0, \sigma_{x}^2),
     y \sim \mathcal{N}(0, \sigma_{y}^2)\\
     v_x \sim \mathcal{N}(0, \sigma_{v_x}^2),
     v_y \sim \mathcal{N}(0, \sigma_{v_y}^2)
\end{array}
\EN
The Bayesian estimate of $\theta$ corresponds to maximizing the posterior density
\EQ \label{Bayes}
\thetahat_{Bayes} = {\rm arg~max_{\theta}}~p(\theta \mid \mathcal{Y}) = {\rm arg~max_{\theta}}~p(\mathcal{Y} \mid \theta)p(\theta)
\EN
or minimizing the corresponding negative log likelihood ratio $L_{Bayes}(\theta)=-log(P(\mathcal{Y} \mid \theta)P(\theta))$. In our Gaussian setting, this yields the following \textit{regularized nonlinear least squares} problem:
\begin{equation} \label{Bayes}
\begin{aligned}
\thetahat_{\text{Bayes}} = \arg\min_\theta \Bigg[ 
\frac{1}{\sigma_R^2} \sum_{i=1}^{N} (R_i - r_i)^2 + 
\frac{1}{\sigma_\Omega^2} \sum_{i=1}^{N} (\Omega_i - \omega_i)^2 + \\
\frac{1}{\sigma_V^2} \sum_{i=1}^{N} (V_i - v_i)^2 + 
\frac{x^2}{\sigma_x^2} + 
\frac{y^2}{\sigma_y^2} + 
\frac{v_x^2}{\sigma_{v_x}^2} + 
\frac{v_y^2}{\sigma_{v_y}^2}
\Bigg]
\end{aligned}
\end{equation}
This can be solved iteratively using the Levenberg-Marquardt method, but the solution is sensitive to initial estimates. 
 
\noindent \textbf{Initial position estimate:} Let ($x_{t_i}, y_{t_i}$) be the target's position estimates w.r.t. $S_i$, which is located at ($x_i, y_i$) w.r.t. $S_1$, then the initial estimate for the target position is
\EQ \label{Initial Position Estimates}
\begin{array}{cc}
x^{initial} = \frac{1}{N}\sum_{i=1}^{N}(x_{t_i}\cos{\phi_i}-y_{t_i}\sin{\phi_i}+x_i)\\
y^{initial} = \frac{1}{N}\sum_{i=1}^{N}(y_{t_i}\cos{\phi_i}+y_{t_i}\sin{\phi_i}+y_i)
\end{array}
\EN


\noindent \textbf{Initial velocity estimate:} 
The initial estimate for the target velocity is taken as
\EQ \label{Initial Velocity Estimates}
\begin{array}{cc}
v_x^{initial} = \frac{1}{N}\sum_{i=1}^{N}(V_{i}\cos{(\phi_{i} + \theta_i)})\\
v_y^{initial} = \frac{1}{N}\sum_{i=1}^{N}(V_{i}\sin{(\phi_{i} + \theta_i)})
\end{array}
\EN

\renewcommand{\thefootnote}{\fnsymbol{footnote}}

\noindent \textbf{Covariance Estimation:} Downstream tasks such as centralized tracking require estimates of posterior covariances as well as posterior means.  Although the \(4 \times 4\) posterior covariance matrix can be estimated analytically by Laplace's method\cite{McKay_Book2002}, we consider a direct numerical approach based on computing the posterior density over a grid for the state \((x, y, v_x, v_y)\), centered around the Bayesian estimate (\(\hat{\theta}_{\text{Bayes}}\)).

Figure~\ref{fig:oneShotPipeline} illustrates the one-shot fusion pipeline for a distributed setup with two radar nodes. 

\subsection{Experimental Results}
We evaluate the one-shot fusion algorithm for the two-node network for the configuration and target trajectories described in Fig.~\ref{ExpSettings}. Due to the absence of exact ground truth for target position and velocity at each frame, we use track-level fusion estimates~\cite{4648063} as a benchmark for our one-shot estimates. Viewing self-calibration as ``training'' of the radar network, and one-shot fusion as ``inference'' based on the trained network, we use different tracks for these two stages.  Thus, for each configuration, we use the straight line trajectory for self-calibration when evaluating one-shot fusion for the random trajectory, and vice versa.

    
\begin{figure}[] 
 \centering 
\includegraphics[scale = 0.4]{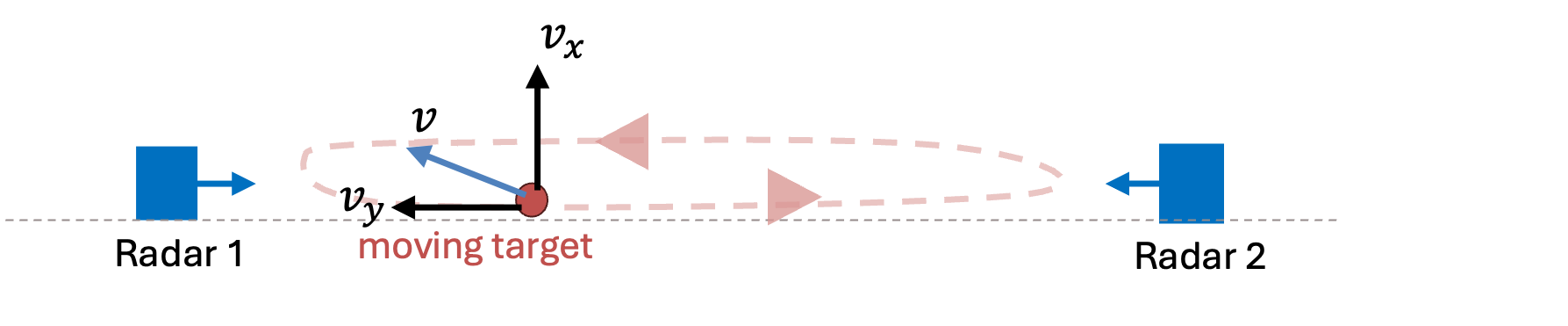} 
\caption{Geometric degeneracy in configuration C}
\label{fig:C_uncertainty}
\vspace{-0.6cm}
\end{figure}
\begin{figure*}[t!]
    \centering
    \subfloat[]{
        \includegraphics[width=0.24\textwidth]{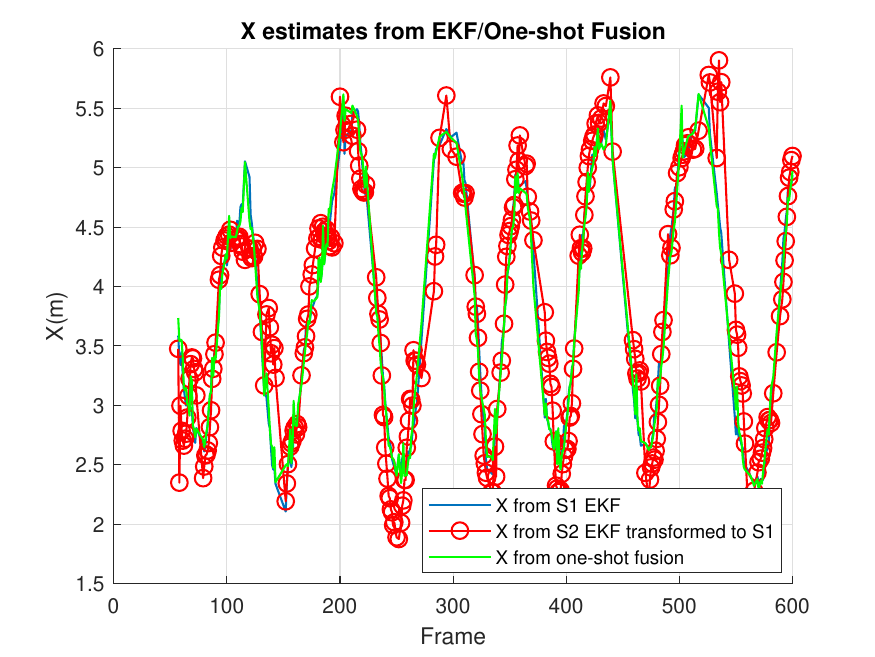}%
        \label{b1}
    }
    \subfloat[]{
        \includegraphics[width=0.24\textwidth]{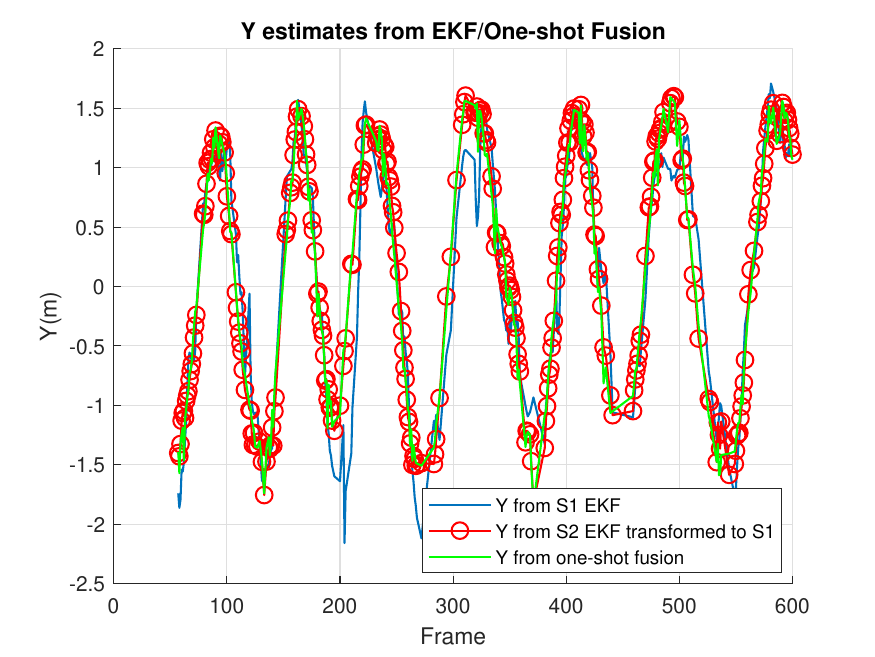}%
        \label{b2}
    }
    \subfloat[]{
        \includegraphics[width=0.24\textwidth]{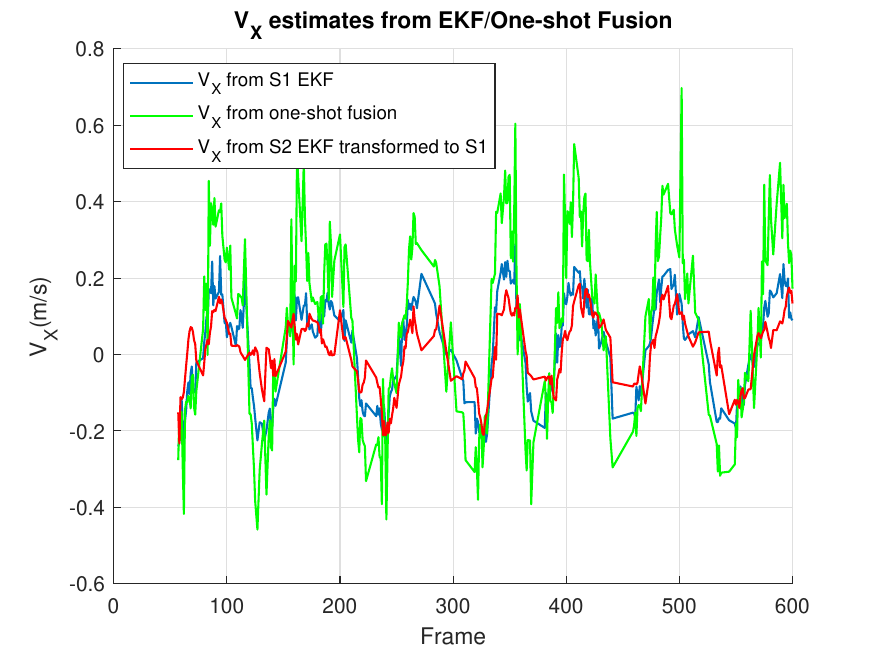}%
        \label{a1}
    }
    \subfloat[]{
        \includegraphics[width=0.24\textwidth]{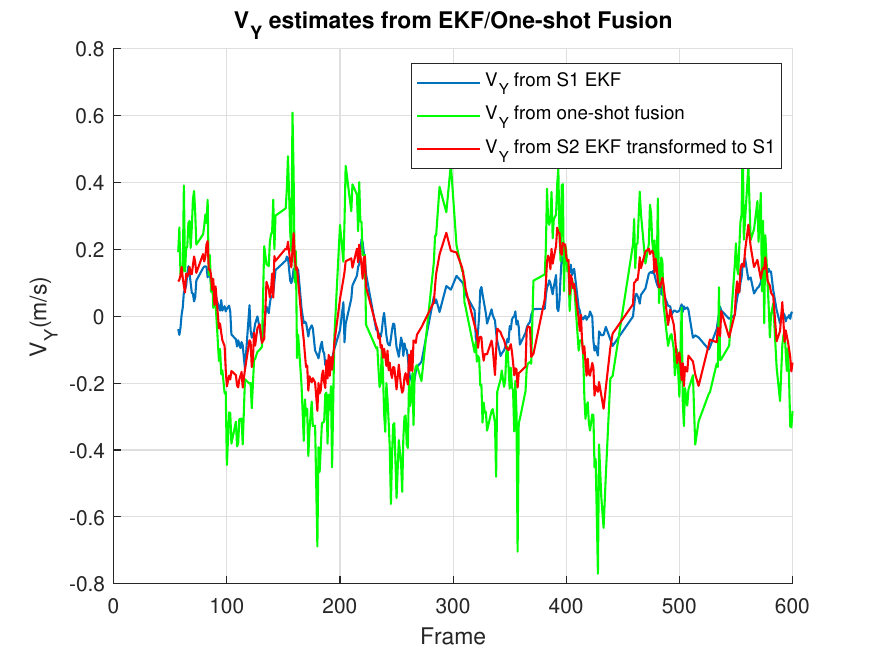}%
        \label{a2}
    }
    \vspace{-0.5em} 
    \caption{$X, Y, V_X, V_Y$ estimates for Setup B random trajectory, after self-calibration using a straight trajectory}
    \label{fig:one_shot_fusion}
\end{figure*}

We set less informative priors (\(\sigma_x = \sigma_y = 3 \, m\)) for position estimates as they are more accurate when using range and angle measurements. In contrast, Doppler measurements only provide estimates of radial velocity. Therefore, we apply stronger priors (\(\sigma_{v_x} = \sigma_{v_y} = 3.5 \, m/s\)) by incorporating our knowledge of average human walking and running speeds. 
For measurement noise, we use \(\sigma_{R} = 0.035 \, m\), \(\sigma_{\Omega} = \pi/4\), and \(\sigma_{V} = 0.1807 \, m/s\), which are the range, spatial frequency, and Doppler resolutions, respectively.  

Geometric degeneracy for one-shot vector velocity estimation occurs whenever the target lies on or near the straight line connecting the two radars, since both radars are then measuring the same radial velocity (with sign flipped).  This corresponds to poorly conditioned maximum likelihood fusion.  While this can happen for any of the three configurations shown in Fig. \ref{ExpSettings}, configuration C, in which the radar arrays face each other, is the most vulnerable, since this happens for a significant portion of the common FoV (Fig. \ref{fig:C_uncertainty}).   Bayesian fusion incorporating realistic priors for target motion addresses such degeneracies and significantly reduces RMSE values, as shown in Table~\ref{tab:oneshot}. 

\begin{figure}[t!]
    \centering
    \subfloat[Region 1 exhibits geometric degeneracy]{
        \includegraphics[scale=0.35]{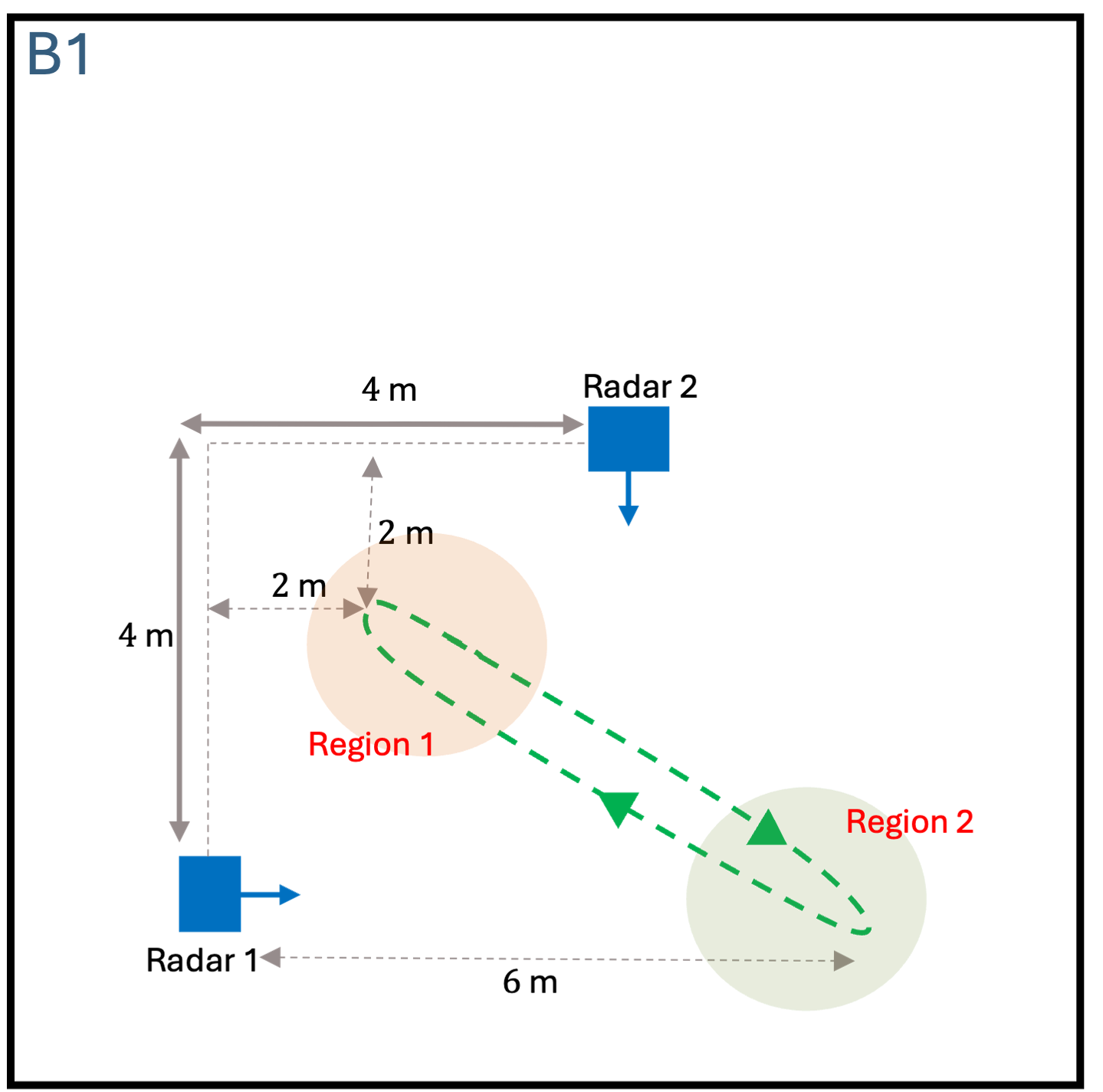}%
        \label{b1_exp_uncertainty}
    }\\
    \subfloat[Region 1 Posterior]{
        \includegraphics[width=0.24\textwidth]{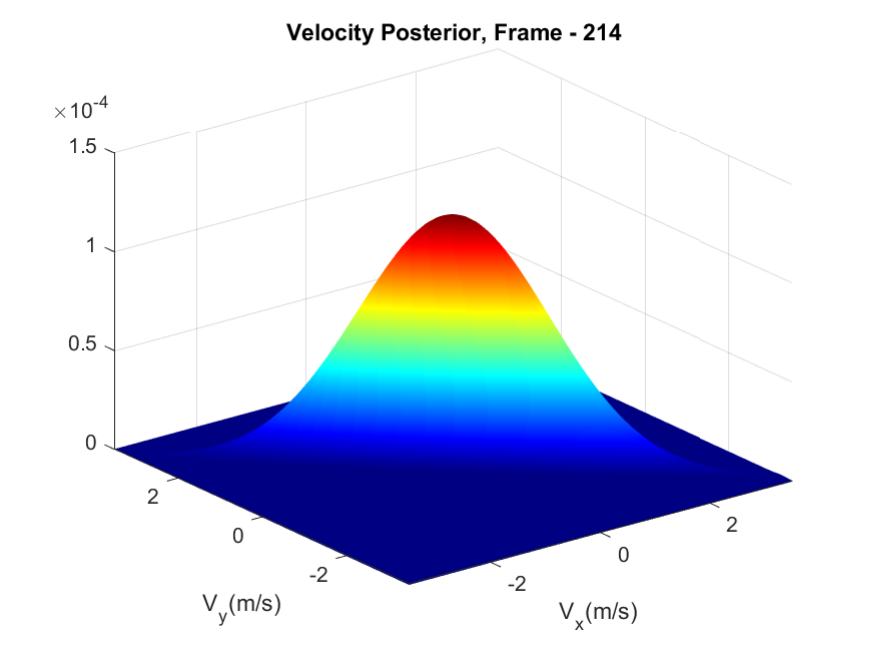}%
        \label{region_1}
    }
    \subfloat[Region 2 Posterior]{
        \includegraphics[width=0.24\textwidth]{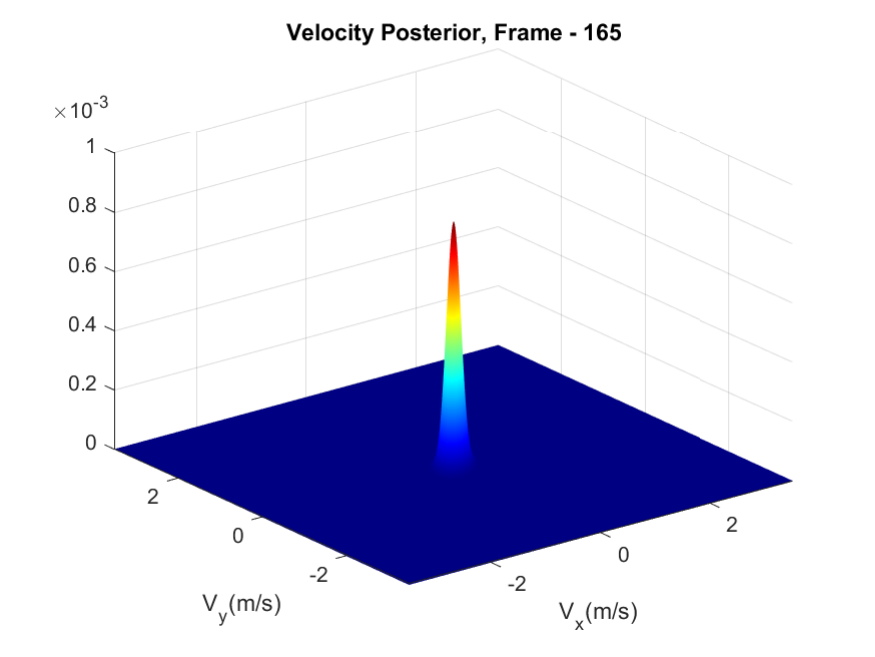}%
        \label{region_2}
    }
    \vspace{-0.5em} 
    \caption{The velocity posterior distribution (b) for a target in region 1 is broader than the posterior (c) for a target in region 2, as expected.}
    \label{fig:uncertainty}
\end{figure}

More generally, the quality of one-shot position and velocity estimates depends on the FoV diversity provided by collaborating nodes. Ideally, the proportion of the common FoV containing targets approximately along the line between radars (leading to velocity estimate degeneracy) or at highly correlated angles of arrival (causing poor position estimates) should be small. Among the three configurations, configuration B offers the best FoV diversity, while configuration C offers the least. This is consistent with the position and velocity RMSE values in Table~\ref{tab:oneshot}: configuration B achieves the best estimates, while configuration C yields the poorest velocity estimates. As illustrated in Figure~\ref{fig:uncertainty}, however, geometric degeneracy can still occur in configuration B, reflected in increased uncertainty in the posterior distribution.
To further validate the accuracy of the proposed one-shot fusion approach, we provide a detailed comparison of the position \((X, Y)\) and velocity estimates \((V_X, V_Y)\) for a random trajectory using configuration B, where self-calibration is performed with the straight trajectory. The estimates are obtained from: 1) the Extended Kalman Filter (EKF) at \(S_1\), 2) the EKF at \(S_2\) transformed into the reference frame of \(S_1\), and 3) one-shot fusion, as shown in Fig.~\ref{fig:one_shot_fusion}. We observe that the EKF estimates from both \(S_1\) and \(S_2\) align closely, confirming the effectiveness of self-calibration. The one-shot velocity estimates exhibit more variations than the smoothed estimates provided by the EKF, but clearly follow the same trends. 
The state covariance matrix, derived from the posterior distribution, allows us to assess the uncertainties tied to the state estimates. For example, in Fig.~\ref{fig:uncertainty}, we display the uncertainties in velocity across two regions using the velocity posterior distribution. 

\begin{table}[t!]
  \centering
  \renewcommand{\arraystretch}{1.3}

  \resizebox{\columnwidth}{!}{%
    \begin{tabular}{@{}l S S c S S c S S@{}}
      \toprule
      & \multicolumn{2}{c}{$A$} & \phantom{abc}& \multicolumn{2}{c}{$B$} &
        \phantom{abc} & \multicolumn{2}{c}{$C$} \\
      \cmidrule(lr){2-3} \cmidrule(lr){5-6} \cmidrule(lr){8-9}
      & {$1$} & {$2$} && {$1$} & {$2$} && {$1$} & {$2$}\\
      \midrule
      \multicolumn{1}{l}{\textbf{Position RMSE (m)}}\\
      Bayesian & 0.3396 & 0.3252 && 0.1699 & 0.1438 && 0.3846 & 0.3690 \\
      ML       & 0.3404 & 0.3248 && 0.1701 & 0.1438 && 0.3839 & 0.3691 \\
      \multicolumn{1}{l}{\textbf{Velocity RMSE (m/s)}}\\
      Bayesian & 0.3609 & 0.3205 && 0.3300 & 0.2656 && 0.4139 & 0.4433 \\
      ML       & 0.3639 & 0.3239 && 0.3439 & 0.2667 && {\bfseries 9.7155} & {\bfseries 6.0543} \\
      \bottomrule
    \end{tabular}
  }%

  \caption{RMSE for one-shot fusion (Bayesian vs ML) with track-level fusion estimates as benchmark. Values indicated in bold indicate cases when ML estimates fail}
  \label{tab:oneshot}
\end{table}

\section{Conclusion}\label{sec:conclusion}

In this work, we present and experimentally demonstrate a framework for collaborative inference in a distributed radar network. We show that simplified point target representations are sufficient for key tasks like self-calibration and one-shot fusion for position and velocity estimation. While our results focus on humans, the approach readily generalizes to other moving targets like vehicles via centroid extraction. The self-calibration accuracy is robust to radar configurations and target trajectory types, while Bayesian one-shot fusion effectively handles geometric degeneracies. This framework lays the foundation for collaborative inference in larger, opportunistically deployed sensor networks. Future work will focus on expanding capabilities, including multi-target tracking, ghost target mitigation, and static scene reconstruction.

\bibliographystyle{IEEEtran}
\bibliography{main}
\vspace{12pt}

\end{document}